\documentclass[11pt]{article}
\usepackage{moriond}

\def\be{\begin{equation}}
\def\ee{\end{equation}}
\def\bea{\begin{eqnarray}}
\def\eea{\end{eqnarray}}

\newcommand{\Photo}{}

\begin{document}
\vspace*{4cm}
\title{Ultra-high energy neutrinos and $W^{\prime}$, $Z^{\prime}$ gauge bosons at the Pierre Auger Observatory}

\author{F. LYONNET}

\address{LPSC, Universit\'e Grenoble-Alpes, CNRS/IN2P3,\\
53 avenue des Martyrs, 38026 Grenoble, France}

\maketitle
\abstracts{
A wide range of models beyond the Standard Model predict
 charged and neutral resonances, generically called $W'$- and $Z'$-bosons, respectively. 
We present a study of the impact of such resonances 
on the deep inelastic scattering of ultra-high energy neutrinos as well as on the resonant charged current $\bar\nu_e e^-$ 
scattering (Glashow
resonance). 
We find that  the effects of such resonances can not be observed with the Pierre Auger Observatory or any foreseeable upgrade of it.
}
\section{Introduction}
New charged and neutral resonances are predicted in many well-motivated extensions of the
Standard Model (SM) such as theories of grand unification (GUTs) or models
with extra spatial dimensions.
These extensions generally do not predict the precise energy scale at which the new heavy
states should manifest themselves. However, for various theoretical reasons
(e.g.\ the hierarchy problem) new physics is expected to appear at the TeV scale and is
searched for at the Large Hadron Collider (LHC) which will soon operate at a center-of-mass
energy of $\sqrt{s}= 13$ TeV.
At the same time, important restrictions on new physics scenarios are imposed by low-energy
precision observables.
On the other hand, highly energetic interactions of cosmic rays in the atmosphere involve processes
at higher center-of-mass energies than those reached by the LHC.
Motivated by this fact, we study the prospects to observe new spin-1 resonances in collisions
of ultra-high energy (UHE) neutrinos with nuclei in the atmosphere as analyzed by the
Pierre Auger Collaboration or a future neutrino telescope.
For example, for neutrinos with an energy of about 10$^{19}$ eV, 
the center-of-mass energy of the neutrino-nucleon interactions is about $\sqrt{s} \simeq 140$ TeV,
considerably extending the energy range accessible at the LHC.
So far, no UHE neutrino events have been observed by the Pierre Auger Observatory
which has led to improved limits on the diffuse flux of UHE neutrinos
in the energy range $E_\nu \ge 10^{18}$ eV \cite{Abreu:2013ppa}.

The potential of the Pierre Auger Observatory for testing new physics scenarios like extra dimensions or the formation 
of micro-black holes has already been studied \cite{Arsene:2013ria}. 
Here, we revisit the predictions for cross sections 
in the SM, and explore the
impact of new charged ($W'$) and neutral ($Z'$) gauge bosons 
on these quantities \cite{Jezo:2014kla}. We address the following questions:
\begin{enumerate}
\item Assuming the LHC does observe new charged or neutral spin-1 resonances,
how would this affect the predicted neutrino cross sections?
\item Assuming the LHC does not discover any new spin-1 resonances,
what are the prospects to observe heavy $W'$- and $Z'$-bosons with masses larger
than 5 TeV using UHE cosmic neutrino events?
\end{enumerate}

\section{$W'$ and $Z'$ gauge bosons}
For definiteness, we consider $W'$ and $Z'$ bosons due to an extended 
$\mathrm{G}(221) \equiv \mathrm{SU}(2)_{1}\times \mathrm{SU}(2)_{2}\times \mathrm{U}(1)_{X}$ gauge group.
Several well-known models emerge naturally from different ways of
breaking the $\mathrm{G}(221)$ symmetry down to the SM gauge group, in particular
Left-Right (LR), Un-Unified (UU), Non Universal (NU), Lepto-Phobic (LP), Hadro-Phobic (HP) and Fermio-Phobic (FP) models.
In this framework, the collider phenomenology has been studied \cite{Abe:2012fb,Jinaru:2013eya} and constraints on the parameter space from low-energy
precision observables derived. In average $Z'$ with masses smaller than 2 TeV are excluded with the most optimistic exclusion limit reaching 3.6 TeV in the UU model. $W'$ bosons are much less constrained: everything above 1 TeV is allowed except in the UU and NU models where the limit is at 2.5 and 3.6 TeV respectively \cite{Hsieh:2010zr}. The collider constrainsts available come from the LHC operated at $\sqrt{s}=7$ TeV and Tevatron and are less stringent or of the same order than the indirect ones \cite{Cao:2012ng}.
In addition, we present results for the Sequential Standard Model
(SSM), where the $W'$- and $Z'$-bosons are just
heavy copies of the $W$- and $Z$-bosons
in the SM.
This is motivated by the fact that the SSM often serves as a benchmark model in the 
literature.
\section{Interactions of UHE neutrinos in the atmosphere}

In the following, we focus on the dominant cross sections of neutrino--nucleon DIS as well as Glashow resonance:

\begin{enumerate}
\item Charged current deep-inelastic scattering (CC DIS): 
$\nu_\ell + N \to \ell^- + X$, $\bar\nu_\ell + N \to \ell^+ + X$.
\item Neutral current deep-inelastic scattering (NC DIS): 
$\nu_\ell + N \to \nu_\ell + X$, $\bar\nu_\ell + N \to \bar \nu_\ell + X$.
\item The Glashow resonance (GR): 
$\bar\nu_e + e^- \to \bar\nu_\ell + \ell^{-}$, $\bar\nu_e + e^- \to q + \bar q'$,
where $q=u,d,s,c,b$.
Obviously, charged current resonant $s$-channel scattering occurs only for incoming 
anti-electron neutrinos.
\end{enumerate}
Note that we have omitted the contributions from non-resonant neutrino--electron scattering which are smaller
by several orders of magnitude. The $W'$ and $Z'$ resonances contribute to the $\nu N$ DIS
with the main contribution comming from the interference with the SM amplitudes. While the GR is entirely negligible at energies $E_\nu \ge 10^8$ GeV there is a new, potentially interesting, resonance 
due to the $W'$-boson which we call GR${}^{\prime}$.

\section{Numerical results}
We now discuss numerical results for the cross sections of 
UHE neutrino interactions in the atmosphere.
For the CC and NC DIS, we consider an isoscalar target and neglect nuclear effects
so that the structure functions are given by the average of the proton and the neutron 
structure functions. 
As is well-known, the UHE neutrino cross sections in DIS are sensitive to the PDFs at very small momentum fractions $x$
down to $x \simeq 10^{-12}$ which results in large uncertainties as shown in Sarkar et al.\cite{CooperSarkar:2011pa}.
On the other hand, the UHE neutrino cross sections are quite insensitive to the lower bound for the integration over the momentum transfer which we set to $1$ GeV$^2$.
In our calculations we use the next-to-leading order (NLO) ZEUS2002\_TR proton PDFs. For simplicity, we neglect the contributions from the NLO Wilson coefficients which
are known to be small. Note that the uncertainties due to the extrapolation of the PDFs into the small-$x$ region and the scale uncertainties are much larger.

\begin{figure}[!h]
	\begin{center}
		\includegraphics[width=0.47\textwidth]{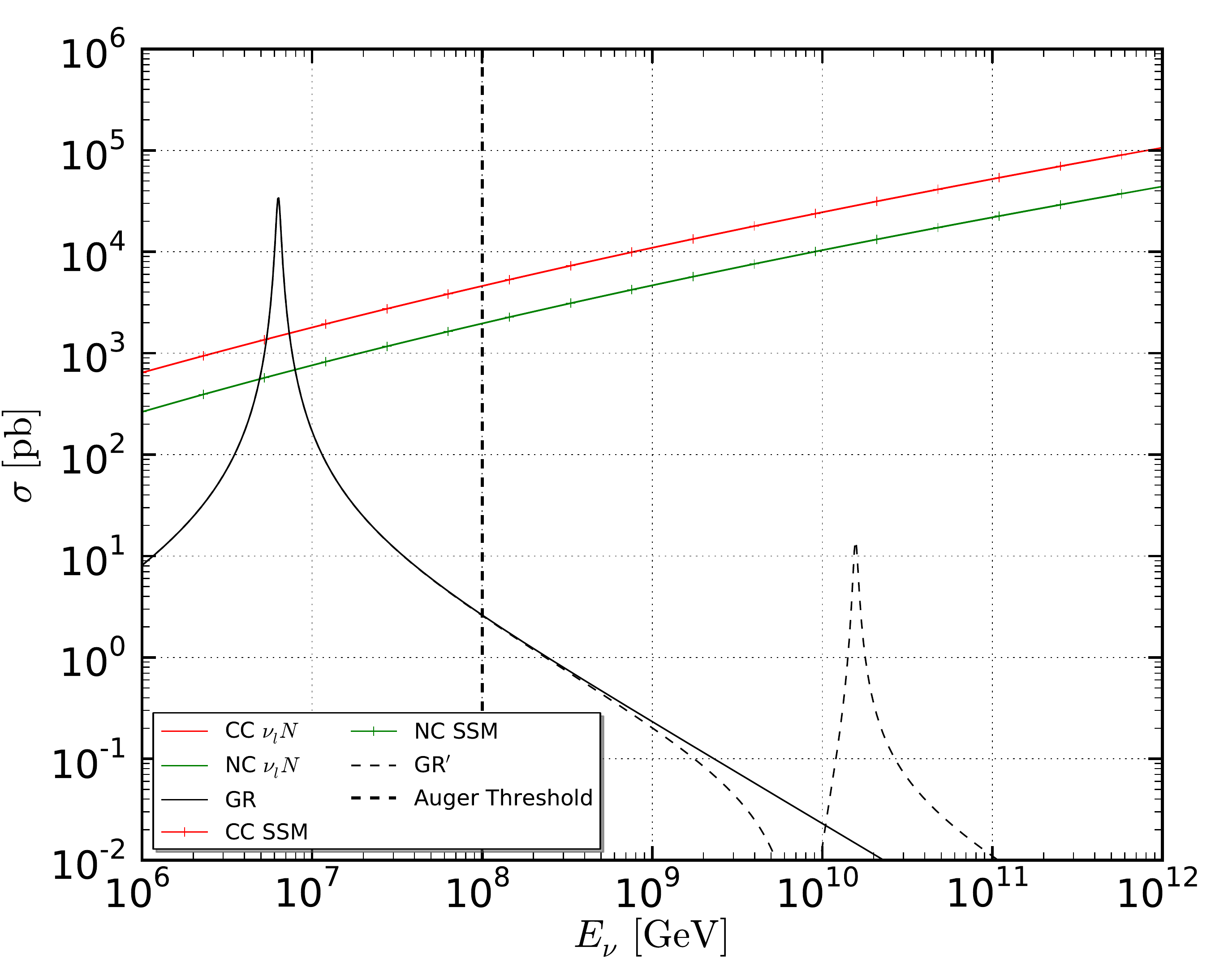}
	\end{center}\vspace{-0.5cm}
	\caption{Total cross sections for CC $\nu_\mu N$ DIS (red line), NC $\nu_\mu N$ DIS (green line) and the Glashow resonance (solid black line)
	in dependence of the incoming neutrino energy. The vertical line at $E_\nu = 10^8$ GeV indicates the lower energy threshold of the Auger Observatory.
The red and green crosses show the CC DIS and NC DIS cross sections, respectively, in the SSM with $M_{W'} = M_{Z'} = 4$ TeV.
The resonant $\bar\nu_e e^-$ scattering including the contribution from the $W'$ resonance is represented by the dashed, black line.}
\label{fig:cross-sections}
\end{figure} 

\begin{figure}[!h]
	\begin{center}
		\includegraphics[width=0.47\textwidth]{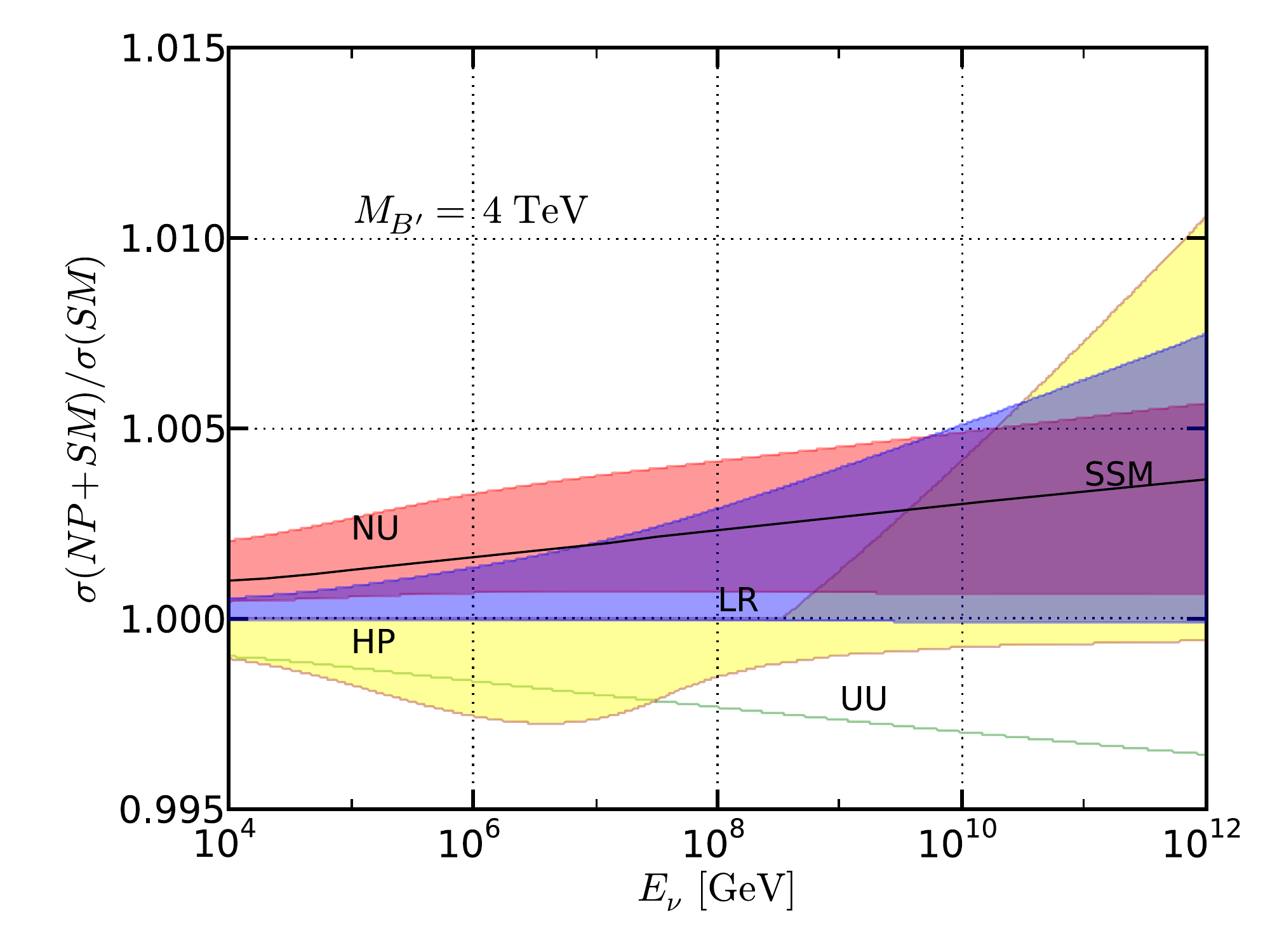}		
	\end{center}\vspace{-0.5cm}
	\caption{The CC+NC $\nu_\mu N$ DIS cross sections in different $\mathrm{G}(221)$ models scaled to the cross section in the SM. The areas have been obtained by fixing either $M_{W'}=4$ TeV or $M_{Z'}=4$ TeV and scanning over the allowed parameter range of the model, see text. For comparison we also show the ratio obtained for the SSM using $M_{W'}=M_{Z'}=4$ TeV.}
\label{fig:ratio}
\end{figure} 

Our total cross sections for CC and NC DIS  are displayed in Fig.~\ref{fig:cross-sections} as a function of the incoming neutrino energy $E_{\nu}$. 
We have verified that our cross section prediction for CC DIS (red line) agrees with the results by Cooper-Sarkar et al.\cite{CooperSarkar:2011pa} 
within a few percent in the entire energy range shown. 
In addition to the SM results, we present predictions for the total cross sections in the SSM (red and green crosses) 
assuming $M_{W'}=M_{Z'}=4$ TeV.
The DIS cross sections in the SM and the SSM differ at the 1\% level and the corresponding curves lie on top of 
each other. Similar observations hold for the other $\mathrm{G}(221)$ models introduced above.
This can be seen in Fig.~\ref{fig:ratio}, where the ratio of the DIS cross sections in the new physics scenario and in the SM is presented.
The areas have been obtained by fixing, depending on the model, either $M_{W'}=4$ TeV or $M_{Z'}=4$ TeV and by
scanning over the allowed parameter spaces of the different models (details have been explained elsewhere \cite{Jezo:2012rm}).
We find that the new physics contributions modify the SM results by at most 1\%, which is much smaller than the theoretical uncertainty
of the DIS cross sections. Similar results hold for masses of the heavy resonance of 5 and 6 TeV.
In Fig.~\ref{fig:cross-sections}, we also show numerical results for the production of hadrons
in resonant $\bar\nu_e e^-$ scattering in the SM (solid, black line) and in the SSM (dashed, black line). 
More specifically, we include the contributions with first and second generation quarks in the final state.
As can be seen, the GR cross section is more than one order of magnitude larger than the total CC neutrino DIS cross section 
at the resonance energy $E_\nu = 6.2 \cdot 10^6$ GeV. 
However, it decreases sharply away from the resonance, and the GR cross section is smaller than the 
CC DIS cross section by several orders of magnitude for energies greater than the Auger Observatory threshold, i.e. $E_{\nu} > 10^8$ GeV.
On the other hand, the contribution from the $W'$ resonance interferes destructively with the SM amplitude 
at energies below $10^{10}$ GeV but leads to a clear enhancement of the cross section in a 
bin around the $W'$-resonance energy $E_{\nu}^{\rm res}=M_{W'}^2/(2 m_e) \simeq 1.56 \cdot 10^{10}$ GeV.
Still it remains more than two orders of magnitude smaller than the DIS cross sections. 
For this reason, the effect of the GR${}^{\prime}$ resonance is irrelevant for events with hadronic showers. Furthermore, even though the GR${}^{\prime}$ can enhance the SM muon production cross section by 7\% at the resonance peak, this is still too small with respect to the uncertainties on the UHE neutrino flux and very small $x$ DIS cross section. Therefore, it seams impossible for general reasons that the very precisely known leptonic cross sections can be used to discover new spin-1 $W'$ and $Z'$ resonances.

\section{Outlook}

We have computed UHE neutrino cross sections in the SSM and $\mathrm{G}(221)$ models including additional
charged and neutral spin-1 resonances.
We find that the effects of such resonances are too small to be observed with the Auger Observatory or any
foreseeable upgrade of it. Conversely, should such resonances be observed at the LHC or a future hadron collider
they will have no measurable impact on the UHE neutrino events. Any deviation from the SM seen in
UHE cosmic neutrino events would require another explanation.

This work is part of a more general ongoing study on the phenomenology of $\mathrm{G}(221)$ models. In this context, we considered the inverse problem at the LHC \cite{Jezo:2012rm} and more recently we developed a tool called PyR@TE \cite{Lyonnet:2013dna} that generates the full two-loop RGEs for arbitrary gauge theories of which the extended gauge group models as the $\mathrm{G}(221)$ are a good example. In the future, this will allow us to address questions such as the vacuum stability in these extensions at the two-loop level.

\section*{References}

\end{document}